\documentstyle[aps,epsfig,amsfonts]{revtex}

\begin{document}

\def\kB{k_B}    
\def\Ref{^{\rm ref}} 
\def\withQ#1{^{(#1)}} 
\def\Q{\withQ Q}
\def\withQo{\withQ{Q_0}}

\def\A{{\cal A}}         
\def\G{{\cal G}}         
\def\C{{\cal C}}         
\def\c{ c_2}             
\def\K{{\cal K}}         
\def\g{\gamma}           
\def\eps{\varepsilon}    
\def\epsbar{\bar\eps}    
\def\u{\tilde u}           
\def\Qinf{Q_\infty}
\def\epsNum{\epsilon_{\tt\#}}

\def\PDE{{\sc pde}}
\def\ODE{{\sc ode}}
\def\HRT{{\sc hrt}}
\def\LOGA{{\sc loga}}
\def\ORPA{{\sc orpa}}
\def\HNC{{\sc hnc}}
\def\RPA{{\sc rpa}}
\def\RG{{\sc rg}}
\def\MC{{\sc mc}}
\def\CPU{{\sc cpu}}

\let\rho\varrho

\title{Implementation of the Hierarchical Reference Theory for simple
one-component fluids}

\author{Albert Reiner$^*$ and Gerhard Kahl}

\address{Institut f\"ur Theoretische Physik and CMS, Technische
Universit\"at
Wien, Wiedner Hauptstr.~8--10, A--1040 Vienna, Austria.\\
{}$^*$e-mail: {\tt areiner@tph.tuwien.ac.at}}

\maketitle

\begin{abstract}
Combining renormalization group theoretical ideas with the integral
equation approach to fluid structure and thermodynamics, the
Hierarchical Reference Theory is known to be successful even in the
vicinity of the critical point and for sub-critical temperatures.  We
here present a software package independent of earlier programs for
the application of this theory to simple fluids composed of particles
interacting via spherically symmetrical pair potentials, restricting
ourselves to hard sphere reference systems. Using the hard-core Yukawa
potential with $z=1.8/\sigma$ for illustration, we discuss our
implementation and the results it yields, paying special attention to
the core condition and emphasizing the decoupling assumption's {\it
r\^ole}.
\end{abstract}


\section{Introduction}

\label{sec:intro}

The {\it Hierarchical Reference Theory} of fluids (\HRT) pioneered by
Parola and Reatto \cite{b:hrt:1,hrt:1,hrt:2,hrt:8,hrt:12,hrt:4} has
been found well capable \cite{hrt:7,hrt:10,hrt:14,hrt:17,hrt:18} of
describing structural and thermodynamic properties of fluids even in
the vicinity of the critical point and for subcritical temperatures,
yielding rigorously flat isotherms in the coexistence region (thus
eliminating the need for Maxwell constructions) and non-classical
values for the critical exponents \cite{hrt:1}. Still, adoption by a
significant part of the liquid physics community of this
renormalization group (\RG) theoretical approach to the integral
equation description of fluids has largely been lacking so far. While
this may be partially attributed to \HRT's inherent difficulties
and rather high computational cost, lack of an easy to use yet
flexible, well-documented implementation of \HRT\ may also have played
a {\it r\^ole}. To fill this gap, we have written a program
\footnote{Available on the world wide web from {\tt
http://purl.oclc.org/NET/ar-hrt-1/}.}
suited as a general framework for the exploration and
application of \HRT\ to simple one-component fluids with hard sphere
reference systems with various combinations of physical systems,
approximations, and solution algorithms.  Within the natural
limitations of the method, it has proved well applicable to a variety
of model systems including the hard-core Yukawa (present contribution)
and hard-core multi-step potentials while most attention has been
devoted to the square well fluid \cite{ar:sw}.

Of course, ours is not the first implementation of \HRT\ for simple
one-component fluids: indeed, there has been a series of earlier
programs \cite{hrt:12,hrt:4,hrt:7} by the authors of the theory and
their collaborators, but it was the one used in \cite{hrt:10,hrt:14},
henceforth referred to as the ``original'' implementation, that was a
vital step in demonstrating the viability of \HRT\ for continuous
systems below the critical temperature; though never published or
formally released, it has been circulating among interested physicists
for quite some time.   Our new software, on the other hand, differs from
its precursors  in
many respects: adoption of a meta language in our version, programming
style and documentation-to-code ratio may be most obvious, number and
nature of hard-coded limitations, important details of the numerical
procedure  and a possible speed
gain through generation of customized code might be less
apparent. Most importantly, though, the original implementation's
structure makes experimentation with different combinations of
approximations, partial differential equation (\PDE)\ solving
algorithms, parameter settings and physical potentials rather
cumbersome; in contrast, the fully modular approach
adoption of a meta language allowed us to take seems far better suited
to a more general survey of \HRT's numerous attractive features. In
addition to the necessary flexibility of our software, great care has
been taken to ensure the numerical soundness of every step in the
calculation and hence of the results produced, so 
that the generation of numerical errors necessarily arising from
finite-precision arithmetic is uniformly spread over all of the
problem's domain. To this end we introduce one central parameter,
$\epsNum$, characterizing the maximum relative error introduced at any
step; together with a number of criteria relying on $\epsNum$, this
parameter governs virtually all of the numerics. Any deviation from
this strategy is made explicit, as are all the other approximations
entering the calculation.  Ultimately, our goal was to provide the
liquid physics community with a general and versatile yet numerically
reliable tool for the systematic exploration and assessment of \HRT\
and of the effects introduced by different approximations.

This paper is meant to serve a twofold purpose: to present the
software we have written and its capabilities, and to provide its
prospective users with some rudimentary documentation. To this end,
after a brief presentation of the standard formulation of \HRT\ for
one-component fluids and some of the theory's properties as far as
they concern our implementation (section~\ref{sec:theory}), we first give
a general outline of our program in section~\ref{sec:global}, only
touching upon the meta-language it has been written in. Due to our
implementation's fully modular design it is only natural to then
proceed by a discussion of the more important of its building blocks
and the various approximations they implement
(sub-sections~\ref{sec:potential} through \ref{sec:solver}). Presentation
and critical assessment
of the kind of results that can be attained with these and concluding
remarks (section~\ref{sec:discussion}) are followed by two appendices
dealing with some technical aspects of the formulation used.

\section{The Theory}

\label{sec:theory}

While a much more detailed account with additional references can be
found in the review article \cite{b:hrt:1}, in what follows we want to
limit ourselves to only a rough sketch of \HRT; in doing so we are
going to stress several aspects --- some of them hardly discussed in
the literature --- relevant to our implementation of \HRT, but only to
the extent necessary for the discussion thereof. No prior knowledge of
\HRT\ is assumed.

The basic ingredient of \HRT, already present in its precursor
\cite{hrt:11}, is the gradual transition from a reference potential
$v\Ref(r)$ to the full potential $v(r) = v\Ref(r) + w(r)$ describing
the interaction between pairs of particles of a fluid, with any one of
the intermediate potentials serving as a reference system with respect
to which the properties of a successor potential are calculated
[superscripts always indicate the system a quantity refers to]. In
our work we have restricted ourselves to the case of a spherically
symmetric pure two-body interaction, and we have taken advantage of the
additional simplifications
possible by identifying the reference system with a pure hard sphere
system, $v\Ref = v^{{\rm hs}}$, as can always be achieved via the
well-known
Weeks-Chandler-Andersen scheme \cite{wca:1,wca:2,wca:3}. Note however
that this restriction to a hard sphere reference system is not present
in the original implementation of \HRT\ \cite{b:hrt:1,hrt:17,hrt:18}; on
the other hand, our program's fully modular framework is flexible
enough to accommodate any of these extensions should the need arise.

Other than \cite{hrt:11}, \HRT\ achieves the transition from $v\Ref$ to
$v$ in infinitesimally small steps. Inspired by momentum-space
renormalization group (\RG) theory, a cut-off wavenumber $Q$ varying
from infinity to zero is introduced, and for every $Q$ the potential
$v\Q = v\Ref + w\Q$ is defined such that Fourier components $k < Q$ of
the perturbational part $w\Q$ of the $Q$-potential $v\Q$ are strongly
suppressed whereas those for $k>Q$ coincide with those of
the original potential $w$. Consequently, the reference system and the
fully interacting system are recovered in the limits $Q\to\infty$ and
$Q\to0$, respectively:
\begin{equation}\label{eq:potential:limits}
\begin{array}{rcl}
v\withQ\infty &=& v\Ref,\\
v\withQ0 &=& v.
\end{array}
\end{equation}
The {\it r\^ole} of the $Q$-potential just introduced becomes clear
when we consider a functional expansion in $\tilde w\Q$ of thermodynamic
and structural properties of the system with pair interaction $v\Q$
[a tilde always denotes the Fourier transform]: as $\tilde w\Q(k<Q)$
is small, the integrals in the expansion are effectively truncated for
$k<Q$, in keeping with the \RG\ picture.

In principle, the precise manner in which the potential is cut off
should not matter, and one can easily conceive of many different ways
of doing so. On the other hand, for such a procedure to be usable it
must not introduce instabilities when truncating the \HRT\ hierarchy,
which is usually done at the two-particle level. Apart from approaches
valid only for special types of potentials, we are aware of only two
cut-off procedures suitable at least for attractive potentials (the
standard formulation of \HRT\ can easily be shown to become unstable
for $\tilde w(0)>0$, as implicitly stated already in \cite{b:hrt:1}); in
our work we opted for the prescription presented in the review article
\cite{b:hrt:1} which seems to have been used almost exclusively so far
\cite{hrt:2,hrt:4,hrt:10,hrt:14,hrt:3} rather than the smooth cut-off
formulation of \cite{hrt:12}, the latter being numerically cumbersome
and predicting non-universal critical exponents. Thus we define the
$Q$-potential $v\Q=v\Ref + w\Q$ by
\begin{equation}\label{eq:cutoff}
\tilde w\Q(k) = \left\{\begin{array}{ccc}
\tilde w(k)&:& k>Q \\
0        &:& k < Q\,;
\end{array}\right.
\end{equation}
in $r$-space, $w\Q$
differs from $w$ by the addition of a convolution integral, {\it viz.{}}
\begin{displaymath}
w\Q(r) = w(r) - {1\over\pi r}
\int_0^\infty\left(
{\sin Q\,(r'-r)\over r'-r} - {\sin Q\,(r'+r)\over r'+r}
\right)\,r'\,w(r')\,{{\rm d}} r'.
\end{displaymath}
Obviously this $Q$-potential is a rather artificial function in
$r$-space hardly resembling the full potential except in the limits of
eq.~(\ref{eq:potential:limits}); furthermore,  the range in $r$-space over
which $v\Q$ has to be considered is bound to be much larger than that
of the original potential --- a property immediately carrying over to
related quantities, the direct correlation functions in particular. As
an immediate consequence, numerical Fourier transformations involving
the $Q$-potential or any of the correlation functions for the
$Q$-system are computationally expensive and must be treated with
extreme care; in fact, they should be avoided if possible at all, with
obvious repercussions for the implementation of the core condition
({\it v.~i.{}}).

In the transition from $v\Ref$ to $v$ or, equivalently, from
$Q=\infty$ to $Q=0$, \HRT\ treats the $Q$-system as reference for the
properties of the system with infinitesimally lower cut-off $Q-{{\rm d}}
Q$; re-summation of terms in the resulting expansions in
${{\rm d}} Q$ and identification of quantities with a well-defined limit
for $Q\to0$ finally yields the \HRT\ equations \cite{b:hrt:1,hrt:2}:
for every density $\rho$ there is a formally exact hierarchy of
coupled integro-differential equations involving a suitably modified
free energy $\A\Q$ defined as
\begin{equation}\label{eq:0loop:A}
{\beta\A\Q\over V} = {\beta A\Q\over V}
- {\rho^2\over2}\left(\tilde\phi(0)-\tilde\phi\Q(0)\right)
+ {\rho\over2}\left(\phi(0)-\phi\Q(0)\right)
\end{equation}
($\phi=-\beta\,w$, $\beta=1/\kB\,T$; analogously,
$\phi\Q=-\beta\,w\Q$), a modified two-particle direct correlation
function
\begin{equation}\label{eq:0loop:C}
\C\Q(r) = c_2\Q(r) + \phi(r) - \phi\Q(r),
\end{equation}
and all higher order correlation functions $c\Q_n(r)$, $n>2$. The
additional terms introduced in $\A\Q$ and $\C\Q$ explicitly take into
account a discontinuity at $Q=0$ present in the unmodified free energy
$A\Q$ and direct pair correlation function $c_2\Q$; as is apparent
from eqs.~(\ref{eq:cutoff}), (\ref{eq:0loop:A}) and (\ref{eq:0loop:C}),
modified and
unmodified quantities coincide for $Q=0$. --- Furthermore it should be
noted that it is customary to include the ideal gas terms in the
definition of the $c\Q_n$: for the two-particle case this is an
additional term of $-\delta(\vec r)/\rho$ so that the Ornstein-Zernike
\cite{oz} equation takes the form
\begin{equation}\label{eq:oz}
\tilde c_2=-(1/\rho)-\rho\,\tilde h\,\tilde c_2\,,
\end{equation}
where $h(r)$ is the usual two-particle total correlation function; for
the higher order correlation functions cf.~\cite{b:hrt:1}.

The full hierarchy the derivation of which we just touched upon yields
expressions for the total derivatives with respect to $Q$ for $\C\Q$
and all the $c\Q_n$, $n>2$; for the evolution of $\A\Q$ we have the
particularly simple relation
\begin{equation}\label{eq:dQA}
{{{\rm d}}\over{{\rm d}} Q}\left({\beta\A\Q\over V}\right)
=  {Q^2\over4\pi^2}\ln\left(1-{\tilde\phi(Q)\over\tilde\C\Q(Q)}\right)\,.
\end{equation}
As ${{\rm d}} c_n\Q/{{\rm d}} Q$, $n\ge2$, involves $\A\Q$, $\C\Q$ and all
higher order correlation functions up to $c_{n+2}\Q$ (so that, in
particular, ${{\rm d}}\C\Q/{{\rm d}} Q$ depends on $c_3\Q$ and $c_4\Q$ via
eq.~(\ref{eq:0loop:C})), the equations never decouple and we have to
introduce some kind of closure. In doing so, it is usually desirable
to retain  thermodynamic consistency as embodied in the sum rule
\begin{equation}\label{eq:consistency}
\tilde\C\Q(0) = - {\partial^2\over\partial\rho^2}\left({\beta\A\Q\over
V}\right)
\end{equation}
rigorously true for the exact solution of the full hierarchy; the
derivatives with respect to $\rho$ present in eq.~(\ref{eq:consistency})
then mandate the transition from equations at fixed $\rho$ to a \PDE\
in the $(Q,\rho)$-plane with boundary conditions supplied at two
densities, ${\rho_{\rm min}}$ and ${\rho_{\rm max}}$.  In addition, we need
to retain the core condition,
{\it viz.{}}\ $g(r)=0$ for $r<\sigma$ where $g(r)=h(r)+1$ is the pair
distribution function; indeed it is one of \HRT's main advantages to
conserve information on all length scales, ranging from the
hard-sphere diameter $\sigma(\rho)$ of the reference system at density
$\rho$ and the associated core condition up to the cut-off wave length
$1/Q$, in the limit $Q\to0$ allowing criticality to arise from
fluctuations of arbitrarily large wave length.

As noted above, the long-ranged nature of $w\Q$ and the correlation
functions due to the cutting-off of eq.~(\ref{eq:cutoff}) is a strong
argument in favor of any closure allowing an approximate
implementation of the core-condition without the need for costly
Fourier transforms. This is a likely reason for the up to now
seemingly exclusive use of a closure in the spirit of the {\it
Lowest-Order $\gamma$-ordered Approximation} (\LOGA, \cite{loga1,loga2}) or
the equivalent {\it Optimized Random-Phase
Approximation} (\ORPA, \cite{orpa}) despite this closure's known
deficiencies \cite{hrt:7,hrt:14}: with the argument $\rho$ silently to
be added in earlier equations when used within the context of the
\PDE, we make the {\it ansatz}
\begin{equation}\label{eq:closure}
\begin{array}{rl}
\tilde\C\Q(k,\rho)&{} = \tilde\phi(k,\rho)
+ \g\Q_0(\rho)\,\u_0(k,\rho)
+ \tilde\K\Q(k,\rho)\,,\\
\tilde\K\Q(k,\rho)&{} = \tilde\G\Q(k,\rho) + \tilde\c\Ref(k,\rho)\,,\\
\tilde\G\Q(k,\rho)&{} = \sum_{n=1}^\infty\g\Q_n(\rho)\,\u_n(k,\rho)\,,
\end{array}
\end{equation}
thereby introducing a set of $Q$-independent basis functions $u_n$ and
corresponding expansion coefficients $\g\Q_n$. Here, $u_0(r,\rho)$ is
chosen proportional to $w(r,\rho)$ (which has the undesirable effect
that, from eqs.~(\ref{eq:closure}) and (\ref{eq:0loop:C}), the correlation
function $c\Q_2$ of the $Q$-system depends upon the full potential $w$
rather than $w\Q$, as would be appropriate) and normalized so that
$\tilde u_0(0,\rho)=1$; the $u_n(r,\rho)$, $n\ge1$, on the other hand,
are taken to form a basis for a suitable function space over
$[0,\sigma(\rho)]$.  With these provisions, the problem of
implementing both core condition and thermodynamic consistency reduces
to that of an appropriate choice of the expansion coefficients
$\g\Q_n(\rho)$, $n\ge0$, for every point in the $(Q,\rho)$-plane. With
the short-hand notations
\begin{displaymath}
\alpha\Q(\rho) = {\partial^3\over\partial Q\partial\rho^2}
\left({\beta\A\Q\over V}\right)
\end{displaymath}
and, for an
arbitrary function $\psi(k,\rho)$,
\begin{equation}\label{eq:c2Int:def}
\hat{\cal I}\Q[\psi(k,\rho),\rho] =
\int_{{\Bbb R}^3} {{{\rm d}}^3k\over(2\pi)^3}\,{\psi(k,\rho)\over
\left(\tilde c\Q_2(k,\rho)\right)^2}\,,
\end{equation}
the
condition (\ref{eq:consistency}) for thermodynamic consistency is easily
re-written as
\begin{equation}\label{eq:consistency:gamma}
{\partial\g\Q_0(\rho)\over\partial Q} =
-\alpha\Q(\rho)
-\sum_{n=1}^\infty
{\partial\g\Q_n(\rho)\over\partial Q}\,\tilde u_n(0,\rho)\,,
\end{equation}
and following \cite{hrt:4} the core condition can be shown to be
equivalent to
\begin{displaymath}
\sum_{n=0}^\infty\hat{\cal I}\Q[\tilde u_j(k,\rho)\tilde u_n(k,\rho),\rho]
\,{\partial\g\Q_n(\rho)\over\partial Q}
= {Q^2\over2\pi^2} 
\,{\tilde\phi(Q,\rho)\,\tilde u_j(Q,\rho)
\over\tilde\C\Q(Q,\rho)
\,\left(\tilde\C\Q(Q,\rho)-\tilde\phi(Q,\rho)\right)}\,,
\qquad j \ge 1.
\end{displaymath}
The latter can be combined with eq.~(\ref{eq:consistency:gamma}) to the
more
explicit
\begin{equation}\label{eq:matrix:infty}
\begin{array}{c}
\sum_{n=1}^\infty
\hat{\cal I}\Q\left[\tilde u_j(k,\rho)\,\left(\tilde u_n(k,\rho)
-\tilde u_0(k,\rho)\,\tilde u_n(0,\rho)\right),\rho\right]
\,{\partial\g\Q_n(\rho)\over\partial Q}\\
=
\alpha\Q(\rho)\,
\hat{\cal I}\Q\left[\tilde u_j(k,\rho)\,\tilde u_0(k,\rho),
\rho\right]

+ {Q^2\over2\pi^2}
\,{\tilde\phi(Q,\rho)\,\tilde u_j(Q,\rho)
\over\tilde\C\Q(Q,\rho)
\,\left(\tilde\C\Q(Q,\rho)-\tilde\phi(Q,\rho)\right)}\,,
\quad j\ge1;
\end{array}
\end{equation}
as both the sum rule and the core condition must hold for the
reference system, the above evolution equations for the $\g\Q_n$ are
readily supplemented with the initial conditions $\g\Ref_n = 0$, $n
\ge0$.

As eq.~(\ref{eq:matrix:infty}) stands, it is no more amenable to direct
implementation than the previous formulation; not only must this
infinite-dimensional matrix equation be truncated to a finite number
$1+{N_{\rm cc}}$ of basis functions, but even then the resulting integrals
need to be evaluated at every $Q$ and $\rho$ --- a tedious process no
less demanding than the Fourier transformations this approach is meant
to replace. What makes this closure manageable, however, is the
observation that the discontinuity in $\hat{\cal I}\Q$'s integrand due to
the appearance of $\tilde c\Q_2$ instead of the continuous $\tilde\C\Q$
leads to a term in $\partial \hat{\cal I}\Q[ \psi(k,\rho), \rho] / \partial
Q$ made up of functions evaluated at $k=Q$ alone; following
\cite{hrt:4}, only this single term is retained, leading to the
approximation
\begin{equation}\label{eq:dQc2Int:approx}
{\partial\over\partial Q}\hat{\cal I}\Q[ \psi(k,\rho),\rho]
=
\psi(Q,\rho)\,{Q^2\over2\pi^2}\,
 {2\,\tilde\C\Q(Q,\rho)\,\tilde\phi(Q,\rho)-\left(\tilde\phi(Q,\rho)\right)^2
\over
\left(\tilde\C\Q(Q,\rho)\right)^2\,
\left(\tilde\C\Q(Q,\rho)-\tilde\phi(Q,\rho)\right)^2}
\end{equation}
leaving out the non-local contribution
$-2\*\sum_{n=0}^\infty\hat{\cal I}\Q[ \psi(k,\rho)\,\tilde u_n(k,\rho) /
\tilde
c\Q_2(k,\rho),\rho]\*(\partial\g\Q_n(\rho)/\partial Q)$; with the
above approximation, the task of evaluating one of the integrals of
eq.~(\ref{eq:matrix:infty}) reduces to only an
initial integration for the reference system followed by the solution
of an ordinary differential equation (\ODE) coupled to the \HRT-\PDE\
as well as analogous \ODE s for all the other integrals of the
$\hat{\cal I}$-type. 

Of course, to fully specify the mathematical problem, the
\PDE\ must be amended by both initial and boundary conditions; while
the former take the simple form of vanishing expansion coefficients
({\it v.~s.{}}), the latter also impose some constraint on the $\g\Q_n$.  
However, as long as we
retain the core condition, such an additional constraint is already
sufficient to determine the expansion coefficient $\g\Q_0$; unless the
$\g\Q_0$ so found exactly reproduces eq.~(\ref{eq:consistency:gamma}),
thermodynamic consistency can no longer be imposed without introducing
mathematical inconsistencies. By the same token,
eq.~(\ref{eq:matrix:infty}), derived by incorporating the sum rule
(\ref{eq:consistency})
into the core condition, is no longer valid but must be changed to
\begin{equation}\label{eq:matrix:g0fixed}
\begin{array}{rl}
\sum_{n=1}^\infty
\hat{\cal I}\Q\left[\tilde u_j(k,\rho)\,\tilde u_n(k,\rho),\rho\right]
\,{\partial\g\Q_n(\rho)\over\partial Q}
=
& - \hat{\cal I}\Q\left[\tilde u_j(k,\rho)\,\tilde u_0(k,\rho),
\rho\right]
\,{\partial\g\Q_0(\rho)\over\partial Q}\\
& + {Q^2\over2\pi^2}
\,{\tilde\phi(Q,\rho)\,\tilde u_j(Q,\rho)
\over\tilde\C\Q(Q,\rho)
\,\left(\tilde\C\Q(Q,\rho)-\tilde\phi(Q,\rho)\right)}
\,,\quad j\ge1.
\end{array}
\end{equation}
to reflect the transition from eq.~(\ref{eq:consistency}) to said
constraint
determining the $\g\Q_0$ appearing on the above equation's right hand
side; furthermore, elimination of thermodynamic consistency obviously
means decoupling the \PDE\ to a set of \ODE s at fixed density.  This
is exactly what is needed at the boundaries ${\rho_{\rm min}}$ and
${\rho_{\rm max}}$ of
the density interval considered where the $\rho$-derivatives defining
$\alpha\Q(\rho)$ cannot be evaluated: While specialization to
${\rho_{\rm min}}=0$ uniquely determines the solution there by virtue of
the
divergence of the ideal gas term $-1/\rho$ in $\tilde c\Ref_2$
(cf.~appendix \ref{sec:app:rewrite}), some prescription for finding
$\g\Q_0$, accompanied by the necessary modification of the truncated
matrix equation according to eq.~(\ref{eq:matrix:g0fixed}), must be imposed
at ${\rho_{\rm max}}$; as long as we retain eq.~(\ref{eq:matrix:infty}) in
the \PDE's
domain's interior and numerically evaluate $\alpha\Q(\rho)$ there, it
is natural to use the \LOGA/\ORPA\ condition $\g\Q_0({\rho_{\rm max}})=0$
\cite{hrt:10} at the boundary, which is sufficient to also determine
$\partial\A\Q({\rho_{\rm max}})/\partial Q$ (or, equivalently, appendix
\ref{sec:app:rewrite}'s $f(Q,{\rho_{\rm max}})$) from the
$\g\Q_n({\rho_{\rm max}})$, $n\ge1$,
alone.

Unfortunately, it turns out that a scheme retaining $\alpha\Q(\rho)$
in eq.~(\ref{eq:matrix:infty}) for ${\rho_{\rm min}}<\rho<{\rho_{\rm max}}$
presents
significant numerical problems for all but extremely high
temperatures, precluding reaching $Q=Q_0$ at least for the potentials
that we have looked at. This is where the so-called ``decoupling
assumption'' comes into play: based upon the different ranges of
$u_0(r)$ and $u_n(r)$, $n\ge1$, the authors of \cite{hrt:4} argue that
$\alpha\Q(\rho)=0$ might be a good approximation, thus eliminating the
$\hat{\cal I}$-integral on the right hand side of
eq.~(\ref{eq:matrix:infty}); it
turns out that this change, invariably adopted in all later
publications, is often sufficient to allow generating a solution all
the way to $Q=Q_0$. From our previous discussion of
conditions imposed on the $\g\Q_n$ it should be obvious that this
decoupling assumption is incompatible not only with the \LOGA/\ORPA\
condition $\g\Q_0(\rho)=0$ retained in the original implementation for
$\rho={\rho_{\rm max}}$ but also with thermodynamic consistency
(eq.~(\ref{eq:consistency})) altogether; thus we are left with only a few
possibilities: we may either retain logical consistency by using the
decoupling assumption $\alpha\Q(\rho)=0$ as a closure for the
\HRT-equations, reducing the \PDE\ to a set of \ODE s in $Q$ only; or
we may prefer to retain the structure of a \PDE\ so as to make use of
thermodynamic consistency at least to a certain degree; yet another
possibility is to maintain both mathematical and thermodynamic
consistency by not implementing the core condition at all. The
original implementation's approach relying on three mutually
incompatible concepts, {\it viz.{}}\ the \LOGA/\ORPA\ condition at
${\rho_{\rm max}}$,
decoupling, and the compressibility sum rule, seems particularly
unattractive; at least one should use the decoupling assumption as a
boundary condition at high density instead.

Retaining thermodynamic consistency in the form of
eq.~(\ref{eq:consistency:gamma}) as well as, in an approximate, way, the
core
condition via the truncated eq.~(\ref{eq:matrix:infty}) or
(\ref{eq:matrix:g0fixed}) together with the approximation
(\ref{eq:dQc2Int:approx}),
we thus arrive at a set of equations implementing \HRT\ with the
\LOGA/\ORPA-like closure (\ref{eq:closure}) on the two-particle level well
suited for numerical processing. While these expressions lend
themselves to discretization in a straightforward way, it is
computationally much more convenient to cast the \PDE\ in a form
superficially resembling a quasi-linear one \cite{hrt:10} so that an
implicit finite-difference scheme requires only the inversion of a
tridiagonal matrix. The re-writing we adopted --- detailed in
appendix~\ref{sec:app:rewrite}, very similar to the one of \cite{hrt:10}
---
results in the introduction of an auxiliary function $f(Q,\rho)$ so
that the \PDE\ implied by eqs.~(\ref{eq:dQA}) and (\ref{eq:consistency})
can be
written in the form
\begin{equation}\label{eq:quasilinear}
{\partial\over\partial Q}f(Q,\rho)
= d_{00}[ f,Q,\rho]
+ d_{01}[ f,Q,\rho]\,{\partial\over\partial\rho}f(Q,\rho)
+ d_{02}[ f,Q,\rho]\,{\partial^2\over\partial\rho^2}f(Q,\rho)
\,.
\end{equation}

Now it is easy to demonstrate the \PDE's stiffness for subcritical
temperatures: recalling the definitions of appendix~\ref{sec:app:rewrite},
we find that the inverse isothermal compressibility $1/\kappa_T$ of
the system with potential $v$ can be written
as $1/\kappa_T = -\rho^2\*\tilde w(0,\rho) / \epsbar(0,\rho)$, where
$\epsbar(Q,\rho)+1 = \eps(Q,\rho) \propto \exp(f(Q,\rho)\,\tilde
u_0^2(Q,\rho))$. For subcritical temperatures there is a density
interval $[\rho_v,\rho_l]$ where $1/\kappa_T = 0$ (implying
diverging $\epsbar$ and hence $f$); here, $\rho_v$ and $\rho_l$ are
the densities of the coexisting vapor and liquid, respectively (recall
that \HRT\ yields rigorously flat isotherms within the coexistence
region, with binodal and spinodal coinciding in three dimensions
\cite{hrt:8}). By construction, however, the limit $Q\to0$ is a
continuous one (cf.~eqs.~(\ref{eq:0loop:A}) and (\ref{eq:0loop:C})) so that
$f$,
$\eps$ and related quantities must be large already well before $Q=0$
is reached; at the same time, the \RG\ mechanism introduced by the
definition of $w\Q$ via eq.~(\ref{eq:cutoff}) precludes any divergence at
non-vanishing~$Q$. Considering the region of the $(Q,\rho)$-plane
where $f$ and $\eps$ are large, we easily find from the explicit
expressions of appendix~\ref{sec:app:rewrite}\ that the $d_{0i}$, and
hence $\partial f(Q,\rho)/\partial Q$, are of order $\eps^1$;
restricting ourselves to a specific density $\rho$ and sufficiently
small $Q$ (so that $\tilde u_0(Q,\rho) = 1$; extension of the argument
to a larger $Q$-range is cumbersome but straightforward) we can
write
\begin{displaymath}
{{{\rm d}} f(Q,\rho)\over {{\rm d}} Q} = e^{f(Q,\rho)} d_0(Q)\,,
\end{displaymath}
where $d_0$ is now of order unity. Inspection of the solution of this
\ODE\ immediately shows that the average of $d_0(Q)$ over the interval
$[ Q_1, Q_2]$, $0<Q_1<Q_2$, is rigorously bounded from above by
$\exp(-f(Q_2,\rho)) / (Q_2-Q_1)$ or else there were a singularity in
that $Q$-interval; translating back to $f$ we see that, while
$\vert\partial f(Q,\rho)/\partial Q\vert$ is still of order $\eps$,
$f$ must be a rapidly oscillating function of $Q$ (with a period of
order $\eps^{-1}$), the average slope of which is much smaller, {\it
viz.{}}\
of order $1/Q$.

It should be noted that this stiffness is not an artifact of the
re-formulation of the \PDE\ as summarized in appendix~\ref{sec:app:rewrite}
but is manifest just the same when directly solving the
\PDE\ for the modified free energy $\A\Q(\rho)$ rather than that for
the auxiliary $f(Q,\rho)$ \cite{ar:th}. The above argument relies only
on some general properties of \HRT\ in the current formulation as
applied to one-component fluids: the divergence of the isothermal
compressibility in the coexistence region (the reproduction of which
is one of \HRT's main achievements), continuity of the limit $Q\to0$,
and the suppression of divergences for $Q>0$ as a result of the \RG\
mechanism implemented via the truncated potential $v\Q$. The essential
additional ingredient, {\it viz.{}}\ the behavior of the ratio of the $Q$-
and
the $\rho$-derivatives as the divergence in the compressibility builds
up, while obvious in the formulation via $f$, is not easily seen in
terms of $\A\Q(\rho)$; this, however, comes as no surprise since $f$
is essentially the free energy's derivative with respect to $Q$ so
that we would have to reason about third- and second-order derivatives
rather than first- and second-order ones if we were to repeat the
arguments without resorting to the re-writing of
appendix~\ref{sec:app:rewrite}.

\section{Overview of the program}

\label{sec:global}

With the theory and approximations outlined in the previous section,
we are now in a position to undertake the task of implementing \HRT\
for one-component fluids in fully standards  conforming {\tt Fortran-90};
the only non-standard feature
we make use of is the availability of the special values {\tt NaN} and
$\pm{\tt Inf}$ for numerically undefined values and signed overflows,
respectively, as defined in the floating-point standard
IEEE~754. These requirements should not pose a serious restriction for
our program's possible users: after all, {\tt Fortran-90}\ compilers have
been
available for a wide range of platforms for several years, and the
desired floating-point behavior can usually be requested --- albeit at
a small performance penalty --- via compiler switches. While our
implementation is more appropriately described as a collection of
mutually compatible building blocks rather than as a monolithic
program so that the details of the numerical procedure are best left
to these parts, for the combination of different selections to work
all versions of all the modular constituents must adhere to a common
model of the computation:

Most obviously, we have to make the transition from the \PDE's domain,
{\it viz.{}}\ the infinite strip $[ 0,\infty) \times [{\rho_{\rm min}},
{\rho_{\rm max}}]$,
to a discrete mesh defined by a finite number of points in the
$(Q,\rho)$-plane. Evidently, the placement of these ``nodes'', as we
shall call them, is of utmost importance for the quality of the
discretization so that it is only natural to define $\epsNum$, the
central parameter governing all of the numerics, in terms of the
properties of this mesh: the coarser a mesh we chose, the larger
$\epsNum$ will be. --- In principle, the locations of the nodes, the
data structures of which are organized in linked lists, can be chosen
freely; in particular, the cut-offs $Q$ of all the systems
in such a list of nodes are not taken to necessarily coincide, even
though this is usually the case except for a low-density boundary at
$\rho=0$.  As for
the densities of the nodes, implementation of the core condition via
the truncated eq.~(\ref{eq:matrix:infty}) and eq.~(\ref{eq:dQc2Int:approx})
makes
anything but constant (though not necessarily equispaced) density
values impractical; if the grid is to be refined for low $Q$,
additional nodes must be inserted at the same densities in all the
node lists in the calculation.  --- After initialization of the nodes' data
structures, solution of the \PDE\ proceeds by applying a (possibly
iterated) predictor-corrector scheme to generate an approximate
solution for the nodes most advanced towards $Q=0$ from the
information available through the node lists at higher $Q$; in the
interest of the code's simplicity, the number of such node lists has
been fixed to exactly three, thus facilitating determination of
appropriate step sizes $\Delta Q$. 

As mentioned before, the goals set for our implementation of \HRT\
necessitate a fully modular architecture of our program; and while we
did not want to forgo the well-known advantages of {\tt Fortran-90}\ for
the
numerical work, experience with prior versions of our code taught us
that the kind of flexibility we need cannot be accommodated within the
rather rigid framework {\tt Fortran-90}'s modules with their one-way flow
of
information provide. Instead we opted for a simple
meta-language\footnote{Available on the world wide web from {\tt
http://purl.oclc.org/NET/arfg/}.}
for self-configuring construction of code customized to the chosen
combination of approximations and the physical system at hand, at the
same time enhancing readability and maintainability of the source and
encouraging modularization; for a more detailed discussion of this
approach and the numerous technical advantages it affords we refer the
reader to \cite{ar:th}.

\subsection{Main parts}

\label{sec:mainparts}

Our software can best be understood as a collection of mutually
compatible and freely exchangeable building blocks corresponding to the
underlying physical and mathematical notions; the resulting natural
organization of the code cleanly separating conceptually unrelated
approximations is a direct consequence of our adoption of a
meta-language and the use of automatic code generation techniques. The
implementation's modular constituents, henceforth dubbed ``main
parts'', must, however, be clearly distinguished from {\tt Fortran-90}'s
modules:
in general, there is no simple mapping from main parts to modules, and
every main part may give rise to any number of modules, incorporating
all the information available within the code base.

In the following sub-sections we take a closer look at some of the
main parts, their physical meaning, the algorithms and approximations
implemented, and at some of the information they make available to the
other parts; we will, however, exclude from this discussion the program's
infrastructure, {\it e.~g.{}}\ the implementation of logging, of reading
and
parsing of options files, handling of node lists and the definition of
a versatile, lossless  and storage-efficient
albeit platform dependent file format for the results at $Q=Q_0$. In a
similar vein, we only mention the assortment of accompanying tools for
reading these files and dumping their content in human-readable or
{\tt Mathematica}-usable form, for locating the critical point or
calculating phase diagrams. --- Thus only main parts ``potential'',
``reference'', ``ansatz'' and ``solver'' remain to be discussed:

\subsection{Properties of the potential}

\label{sec:potential}

First and foremost, we obviously have to provide 
information on the fluid's potential $v=v\Ref+w$ and its properties:
this is the purpose of the main part labeled ``potential''.  Just as
the full potential is a sum of a reference part $v\Ref$ and a
perturbational part $w$, the functions and parameters to be provided
by this main part fall into two distinct categories, pertaining to
either $v\Ref$ or $w$; in addition, as the temperature enters the
calculation only as a pre-factor to $w$, {\it viz.{}}\ via
$\phi=-\beta\*w$,
the inverse temperature $\beta$ is also defined here.

As far as the reference system is concerned, restriction to hard
spheres (for the rationale cf.\ section~\ref{sec:theory}) means that only
a function returning the hard sphere diameter $\sigma(\rho)$ and a
flag indicating any deviation of $\sigma(\rho)$ from the unit of
length need to be made available. --- A similar parameter pertaining
to the perturbational part $w$ of the potential, {\it viz.{}}\ a flag
indicating any density-dependence of $w$, also plays an important {\it
r\^ole} in many parts of the program as substantial simplifications
and, in many cases, significant
speed-ups by caching previous results are possible whenever
$\tilde\phi(Q,\rho)$ only depends on $Q$.   In
addition, at every cut-off $Q$ the program must have access to the
Fourier transforms $\tilde w(Q,\rho)$ and $\tilde\phi(Q,\rho)$ as well as
the derivatives $\partial\tilde\phi(Q,\rho)/\partial Q$ and
$\partial^n\left(\tilde\phi(Q,\rho)\right)^m/\partial\rho^n$, whereas
powers of the volume integral, $\tilde\phi(0,\rho)^n$, and their
derivatives $\partial^n\left(\tilde\phi(0,\rho)\right)^m/\partial\rho^n$
obviously do not depend on $Q$ (here, $m$ and $n$ are appropriate
integers known during code construction). 

For the benefit of the \PDE-solving algorithm, this main part also has
to set a parameter $\lambda_{[ v ]}$ related to the maximum relative
curvature of the second $Q$-derivative of $\tilde\phi(Q,\rho)$, defined
in such a way as to coincide with $\lambda$ for the square
well potential $v^{{\rm sw}[-\epsilon,\lambda,\sigma]}$. Indeed, most of
our
efforts so far \cite{ar:sw} have concentrated on this particularly
simple type of potential given~by
\begin{displaymath}
w^{{\rm sw}[-\epsilon,\lambda,\sigma]}(r) = \left\{\begin{array}{ccc}
-\epsilon &:& r<\lambda\,\sigma \\
0        &:& r>\lambda\,\sigma\,,
\end{array}\right.
\end{displaymath}
where the hard sphere diameter $\sigma$ of the reference part and the
strength $\epsilon$ of the potential are usually chosen as units of
length and energy, respectively. --- Another type of potential we have
implemented is a generalized step potential, {\it i.~e.{}}\ a succession of
stretches of constant $w(r)$; more specifically, the perturbational
part $w(r)$ of an $n$-step potential of this type is just a sum of
square-well potentials, $w=\sum_{i=1}^n
w^{{\rm sw}[-\epsilon_i,\lambda_i,\sigma]}$, 
$ \lambda_i < \lambda_{i+1}$, $1\le i<n$; again, we have only considered
the
$\rho$-independent case, while all the potential's parameters should
be assumed functions of $\rho$ when modeling a specific physical
system. ---  We
have also implemented a $\rho$-independent hard-core Yukawa potential
$v^{{\rm hcy}}$,
\begin{displaymath}
w^{{\rm hcy}[-\epsilon_0,-\epsilon,z,\sigma]}(r) =
\left\{\begin{array}{ccc}
-\epsilon_0 &:& r<\sigma \\
-\epsilon\,{\sigma\over r}\, e^{-z\,(r-\sigma)} &:& r>\sigma\,,
\end{array}\right.
\end{displaymath}
where the parameter $\epsilon_0$, the value of $-w$ inside the core,
defaults to $\epsilon = -w^{{\rm hcy}}(\sigma+)$, which again is usually
chosen to coincide with the unit of energy; any mismatch between
$\epsilon_0$ and $\epsilon$ dominates $\tilde w^{{\rm hcy}}(k)$ for large
$k$
and is found to render unstable at least the numerics.

Regarding the stability of the \PDE, recall from section~\ref{sec:theory}
that an attractive potential (so that $\tilde w(0,\rho)<0$) is a
necessary though not sufficient condition for the stability of the
\PDE.

\subsection{Hard sphere reference system}

\label{sec:reference}

Due to the specialization of $v\Ref$ to hard spheres, the reference
system enters the expressions of section~\ref{sec:theory} only through the
direct correlation function $c\Ref_2$, implementation of which is the
task set for main part ``reference'': drawing upon the information from
main part
``potential'', only some
initialization code and functions for the evaluation of $\tilde
c\Ref_2(Q,\rho)$ and $\partial\tilde c\Ref_2(Q,\rho)/\partial Q$ have to
be exported.

In our program we have so far included two different versions
implementing the Percus-Yevick \cite{hs:8} approximation and the
Henderson-Grundke \cite{hs:2} description; note that all results reported
here have been obtained
using the Henderson-Grundke $\c\Ref$: in a theory relying on internal
consistency conditions like eq.~(\ref{eq:consistency}) as heavily as \HRT,
the thermodynamic inconsistency present in the Percus-Yevick solution
seems particularly undesirable.

\subsection{Discretization, boundary conditions, and other approximations}

\label{sec:ansatz}

Main part ``ansatz'', where all the approximations on the physical and
mathematical level are combined to jointly define a reasonable
numerical model of \HRT, is at the very core of the \PDE-solving
machinery: for the potential the perturbational and reference parts of
which have been described in the two previous sub-sections, the
\HRT-\PDE\ is discretized and solved according to a given set of
approximations and on the mesh defined by the node-lists served by
main part ``solver'' (sub-section~\ref{sec:solver}). More precisely,
``ansatz'' provides a set of facilities in the form of subroutines
with standardized interfaces implementing the various stages of
the computation, {\it viz.{}}\ initialization of the node lists  at
$Q=Q_\infty$ and solution of the \PDE\ according
to a predictor-corrector full approximation scheme. Note, however,
that the code must accommodate the possibilities of both iterating the
corrector step (which may allow reaching the numerical quality
indicated by $\epsNum$ with somewhat larger step sizes, thus speeding
up the calculation) and of discarding part of the solution should
$\epsNum$-based criteria not be met; to aid ``solver'' in these
decisions, care has to be taken to detect and signal numerical
anomalies. Once a step's results have been accepted, ``ansatz'' may
perform additional manipulations of the data structures; most
importantly, the re-scaling of all quantities affected by
exponentiation of $f$ necessary whenever $f$ is large (cf.~our
discussion of the \PDE's stiffness in section~\ref{sec:theory}) is
adjusted  only when
the last corrector's result has been accepted.

\begin{figure}
\epsfxsize=8.6cm
\epsfbox{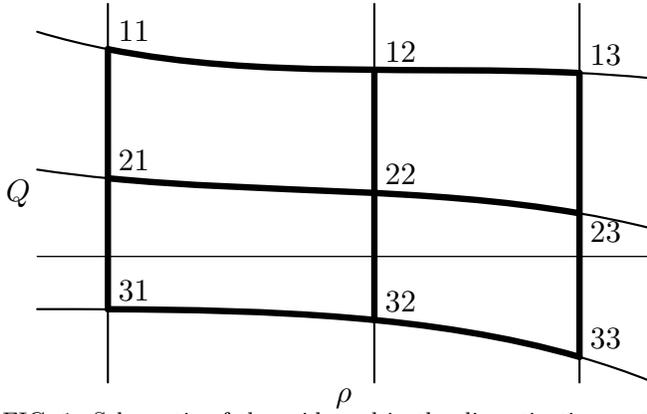}

\caption{Schematic of the grid used in the discretization or the
\PDE\ (cf.\ sub-section~\ref{sec:ansatz}). Assuming use of the
three-point approximation for the second derivatives in the
$\rho$-direction, the discretization is generated from an expansion
around the intersection of the thin horizontal line with the line of
constant density joining the nodes labeled $(i2)$. --- According to
the general model of the computation discussed in section~\ref{sec:global},
a node list's $Q$-values may be $\rho$-dependent, whereas the
$\rho$-values must coincide in all three node lists, though they need
not be equispaced.}
\label{fig:grid}
\end{figure}

Due to the eminent {\it r\^ole} of the consistency condition
eq.~(\ref{eq:consistency}) in constructing a closure to the underlying
\ODE~(\ref{eq:dQA}), the \PDE~(\ref{eq:quasilinear}) for $f(Q,\rho)$ is of
first order in
$Q$ and of second order in $\rho$; assuming the lowest possible number
of nodes in the discretization (extension to higher order is
straightforward), we need at least a $2\times3$ set of
nodes. According to the general model of the computation presented in
section~\ref{sec:global}, however, we instead keep a third node list in
order to allow monitoring of second $Q$-derivatives, so that we use a
discretization on the $3\times3$ grid schematically presented in
fig.~\ref{fig:grid}, including information available via that additional
node list. Locally, the discretization is derived from an expansion
about the midpoint of the nodes labeled $(22)$ and $(32)$ in the
schematic~\ref{fig:grid}, evaluating the second $\rho$-derivative along
the line of constant $Q$ through this point (thin horizontal line in
fig.~\ref{fig:grid}) by estimating the data at the intersection with the
lines of constant density by interpolants defined from node triples
$(i1)$ and $(i3)$, respectively; the resulting finite difference
approximation is applied to every set of three adjacent node-triples,
substituting suitable boundary conditions at ${\rho_{\rm min}}$ and
${\rho_{\rm max}}$.

As indicated in fig.~\ref{fig:grid}, $Q$ is not necessarily constant along
a given node list, whereas the stability of the numerical scheme may
impose certain geometrical constraints regarding the possible
locations of the nodes, {\it e.~g.{}}\ for ensuring that the
Courant-Friedrichs-Lewy criterion \cite{numpde:cfl} is met or for
maintaining convexity of the remaining integration region; a suitable
representation of these constraints is exported and must be taken into
account by main part ``solver''. If the latter decides to insert nodes
at intermediate densities, the code for initializing the inserted data
structures
and for interpolating appropriate quantities is negotiated between the
main parts, depending upon the order of the interpolation formul\ae\
available. A further consequence of having non-constant $Q$ is that
some parts of the density range may reach $Q\approx Q_0$ earlier than
others; in this case, the corresponding nodes are locked, preventing
further modification, and all of the converged nodes except those
necessary for providing a boundary condition for the remaining density
interval are removed from the node lists available to main part
``ansatz''.

In addition to the discretization of the \HRT-\PDE~(\ref{eq:quasilinear})
discussed so far, the implementation of the core condition along the
lines of section~\ref{sec:theory} and appendix~\ref{sec:app:rewrite} is
also
of interest. Relegating discussion of the choice of appropriate basis
function $u_n$, $1\le n\le{N_{\rm cc}}$, to appendix~\ref{sec:app:basis},
we only
point out the extremely slow convergence of the $\hat{\cal I}\Q$-integrals
(\ref{eq:c2Int:def}) that have to be evaluated at $Q=Q_\infty$;
furthermore,
as the integrand is temperature dependent for $k>Q_\infty$, these
integrals have to be evaluated for every isotherm --- a problem that
might be sidestepped by adopting the original implementation's
strategy of consistently using the results for $Q\to\infty$ rather
than those valid at $Q_\infty$ for initialization even though such an
approach introduces a discontinuity at $Q=Q_\infty$. Also, with the
usual choice of $Q_\infty \sim 10^2/\sigma$, integration merely up to
$k=Q_\infty$ can hardly be deemed sufficient; an appropriate upper
integration limit can instead be found by comparing the integrand's
asymptotic behavior with $\epsNum$. --- Main part ``ansatz'' also has
to identify quantities suitable both for monitoring convergence of the
full approximation scheme and for choosing appropriate step sizes
$\Delta Q$ and $\Delta\rho$, and to make available code fragments for
the inspection of nodes in various stages of the computation as well
as a description of the boundary conditions at ${\rho_{\rm min}}$ and
${\rho_{\rm max}}$
including mandatory settings for either of these parameters if
necessary; in particular, most implementations require ${\rho_{\rm min}}=0$
in
order to be able to use the divergence of the ideal-gas term in $\tilde
c\Ref_2$ as a $Q$-independent boundary condition for $f$ (cf.\
appendix~\ref{sec:app:rewrite}).

It is this main part that defines the formulation of \HRT\ and the set of
approximations used in the calculations; of the numerous versions of this
main part we produced only a few are to be mentioned here:  both the
re-implementation of the original
program's approximations for the core and boundary conditions and the
approach combining the \PDE\ with $\alpha\Q=0$ at all densities
including ${\rho_{\rm max}}$, while mathematically inconsistent, retain
thermodynamic consistency at least in some approximate way (cf.~our
discussion of the decoupling assumption in section \ref{sec:theory}); in
addition to these, the two possible approaches at least mathematically
meaningful, {\it viz.{}}\ the \ODE s directly following from decoupling and
the \PDE\ resigning on the core condition for the benefit of the
compressibility sum rule (\ref{eq:consistency}) (with the \LOGA/\ORPA\
prescription $\g\Q_0({\rho_{\rm max}})=0$ as high density boundary
condition)
will also be used in section \ref{sec:discussion}'s presentation of the
results our software yields.

\subsection{Criteria for positioning of nodes}

\label{sec:solver}

If main part ``ansatz'' is to provide a discretization on whatever
mesh is handed to it, it is the task set for ``solver'', the last of
the main parts to be discussed here, to define this very mesh and to
keep track of the numerical solution's quality. Based primarily upon
the value of $\epsNum$  and respecting any
restrictions exported by ``ansatz'', step sizes $\Delta Q$ and
$\Delta\rho$ have to be chosen and checked for compatibility with the
solution generated, iterating or discarding steps if certain criteria
are not met; whenever ``ansatz'' signals an exception 
the last step is discarded, accepting the data in the node list
corresponding to labels $(2i)$ in fig.~\ref{fig:grid} as the best
approximation to the solution for $Q\to0$. At the same time, care has
to be taken to locate and identify any problems in the solution, {\it
i.~e.{}}\
parts of the $(Q,\rho)$-plane where the solution found does not appear
smooth on the scales set by the step sizes, the most basic assumption
underlying any finite-difference calculation; whenever this assumption
no longer holds, the algorithm will react by locally reducing $\Delta
Q$ and $\Delta\rho$, inserting node triples (cf.~sub-section
\ref{sec:ansatz}) in order to achieve the latter. --- Once we find any
nodes
already holding the final results for their respective densities they
must be taken care of as discussed in sub-section \ref{sec:ansatz}; 
integration of the \PDE\ is ended when there is a node with $Q\le Q_0$
for every density in the calculation, or when ``ansatz'' requests an
end either because an error condition has occurred ({\it v.~s.{}}) or
because
the current node list is sufficiently close to $Q=Q_0$ already. --- As
noted in section~\ref{sec:global}, the intimate link between this main
part's task and the numerical quality of the solution generated makes
it natural to here define $\epsNum$, the central parameter governing
the numerics, and it is this part of the program that relies upon
$\epsNum$ and the associated criteria the most; other main parts use
$\epsNum$ for little more than for switching between full analytic
expressions and asymptotic expansions (for a slightly atypical example
of which cf.~appendix~\ref{sec:app:basis}).

Of the two implementations of this main part, one has been written in
the hopes of being able to avoid the problematic region of large
$f(Q,\rho)$ altogether, as is indeed possible for some similar \PDE
s. This implementation makes full use of $\epsNum$, relying on
numerous criteria to control the calculation; in the following
discussion the notation $p^{[x]}_{y}$ refers to customization
parameters that should usually be taken as real numbers of order
unity. A case in point is the choice of the density grid:
even though this is not necessary, we decided to always start with an
equispaced set of $\rho$-values ranging from ${\rho_{\rm min}}$ to
${\rho_{\rm max}}$;
the number $N_\rho+1$ of such density values is related to $\epsNum$
by
\begin{displaymath}
N_\rho = {({\rho_{\rm max}}-{\rho_{\rm
min}})^2\over\epsNum\,p^{[\rho]}_{N_\rho}}\,,
\end{displaymath}
reflecting the importance of second $\rho$-derivatives in the solution
of the \PDE\ as well as the static nature of this set of densities due
to the $\hat{\cal I}$-approach to the core-condition. --- Once $\epsNum$
has
been determined, the system is ready to start determining appropriate
step sizes $\Delta Q$; in particular the assumption that the potential
$v(r,\rho)$ introduces length scales only in the range from
$\sigma(\rho)$ to $\lambda_{[ v]}(\rho)\,\sigma(\rho)$, where
$\lambda_{[ v]}$ is related to the fourth derivative of $\tilde
v(k,\rho)$ with respect to $k$ (cf.~sub-section \ref{sec:potential}),
places an upper bound on the admissible step sizes, {\it viz.{}}
\begin{displaymath}
\Delta Q \le
{\sqrt{12\,\epsNum\,p^{[\Delta Q]}_{\Delta Q_{\rm max}}}
\over\lambda_{[ v]}\,\sigma(\rho)}\,.
\end{displaymath}
On the other hand, for a finite difference scheme to be meaningful at
least a certain number of bits must remain significant in evaluating
the differences, which implies a lower bound on $\Delta Q$
proportional to $Q$, and the solution has to be smooth on the scales
defined by the mesh, which also rules out abrupt changes in the step sizes;
consequently, the ratio of two consecutive $\Delta Q$ steps at the
same density is restricted to lie between $p^{[\Delta Q]}_{\rm
ratio}$ and $1/p^{[\Delta Q]}_{\rm ratio}$. In a similar vein,
considering smoothness in the $\rho$-direction we have to postulate
that $(Q_{(22)} + Q_{(32)})/2$ is greater than either of $Q_{(31)}$
and $Q_{(33)}$, where the labels coincide with those of the nodes of
fig.~\ref{fig:grid}; this condition, unlike the other rules mentioned so
far, does not limit the step sizes $\Delta Q$ at any density $\rho$
but rather determines whether $\Delta\rho$ should be reduced by the
insertion of nodes at an additional density. But the most important
criteria for choosing $\Delta Q$ come from monitoring the solution
generated: for every monitored quantity $x$ we make sure that
\begin{displaymath}
\begin{array}{c}
\sqrt{{1\over\Vert x\Vert_Q}
\,\left\vert{\partial^2x\over\partial Q^2}\right\vert}
\,\Delta Q
\le
\sqrt{\epsNum\,p^{[\Delta Q]}_{x}}\,,
\\
\Vert x\Vert_Q 
=
\max_{k>Q}\vert x(k,\rho)\vert\,;
\end{array}
\end{displaymath}
the quantities taken for $x$ are, of course, defined by ``ansatz'',
and a usual choice is $x(Q,\rho) \in
\left\{f(Q,\rho),1/\tilde\K\Q(Q,\rho)\right\}$. A different set of
quantities $y$, also chosen by ``ansatz'' and usually comprising just
$y(Q,\rho)=f(Q,\rho)$, is used to monitor the convergence of the
predictor-corrector scheme and to determine whether or not the
corrector should be iterated: denoting the absolute difference of
consecutive approximations of $y$ divided by $\Vert y\Vert_Q$ by
$\Delta y$, iterations are performed until $\Delta y <
\epsNum\,p^{[\rm conv]}_{y}$, and the ratio of two consecutive step
sizes is bounded from above by $(\epsNum\,p^{[\Delta Q]}_{y}) ^
{1/\left(2+p^{[\Delta Q]}_{N_{\rm it}}\right)} /
\sqrt{\Delta^{(1)}y}$, where $\Delta^{(1)}y$ is $\Delta y$ evaluated
after the first corrector step. According to simple heuristic
arguments regarding the convergence of corrector iterations and
ignoring the effect of other criteria, an average of $p^{[\Delta
Q]}_{N_{\rm it}}$ calls of the corrector can be expected to solve the
difference equations to within $\epsNum$, and a setting of
$p^{[\Delta Q]}_{N_{\rm it}}>1$ may significantly speed up the
calculation by allowing larger steps to be taken without loss of
accuracy. --- After finding and tentatively using a candidate $\Delta
Q$, we still have to check that the assumptions leading to that
particular choice for $\Delta Q$ actually hold; to this end we
re-evaluate all the criteria with the obvious exception of the one
involving $\Delta^{(1)}y$ after the predictor and discard the step
unless a slightly smaller step size, {\it viz.{}}\ $\Delta Q\,p^{[\Delta
Q]}_{\rm discard}$ ($p^{[\Delta Q]}_{\rm discard}<1$), passes the
tests. If no step size can be found satisfying all the constraints,
the calculation is terminated.

While the above set of prescriptions for finding suitable node
locations has proved indispensible in understanding the behavior of
the \PDE's solution, the oscillatory nature of $f(Q,\rho)$ invariably
linked to the build-up of the isothermal compressibility's divergence
for sub-critical temperatures prevent its use for $\beta>\beta_c$:
considering even the modest value $f\sim10^3$, $\sigma\,\Delta Q$
would have to be smaller than $e^{-10^3}\sim10^{-430}$, which is
obviously completely useless for any practical implementation. Thus,
even though it means loosing control over the level of accuracy in the
solution, we have also implemented a version of this main part with
predetermined step sizes that just happen to often be sufficient for
reaching $Q=Q_0$ even well below the critical temperature while
reproducing the overflow necessary for $\kappa_T$'s divergence in a
density interval the edges of which may then be identified with the
coexisting phases' densities $\rho_v$ and $\rho_l$. Recalling the
behavior of $f$ wherever it is large we obviously have to drastically
reduce $\Delta Q$ as we approach $Q_0$; for this we use the very
prescription introduced by the authors of \cite{hrt:10} and evidently
underlying all later published \HRT\ calculations. 

One last aspect of this main part to be mentioned regards the choice
of $Q_\infty$ in both implementations: As the only reasonable initial
condition for the core condition assumes that the structure at
$Q_\infty$ is basically the same as that for $Q=\infty$ (so that
$c\withQ{Q_\infty}_2 = c\withQ{\infty}_2 = c\Ref_2$ or, equivalently,
$\g\withQ{Q_\infty}_n = \g\withQ{\infty}_n = 0$, $n\ge0$) and the same
set of parameters must also be used for the initialization of nodes at
$Q=Q_\infty+\vert\Delta Q\vert$ (the nodes labled $(1i)$ in
fig.~\ref{fig:grid} in the first step), it is preferable to have
$\partial\g\Q_n(\rho)/\partial Q=0$ at $Q=Q_\infty$; from
eq.~(\ref{eq:matrix:infty}) one immediately concludes that this is
equivalent to
$\tilde w(Q_\infty)=0$ whenever using the decoupling assumption. It is
left to main part ``ansatz'' to decide whether $Q_\infty$ should be
determined in this way, thereby necessarily introducing
$\rho$-dependent $Q_\infty$ when dealing with a $\rho$-dependent
potential.

\section{Discussion and conclusion}

\label{sec:discussion}

In the preceding sections we had to introduce a number of
approximations, some of which may seem rather less justified; their
respective importance for and bearing on our program's predictions of
structure and thermodynamics of simple liquids now remains to be
assessed. As an exact solution with which to compare numerical results
is lacking, for \HRT\ as implemented by our software package to be
considered a reliable tool well applicable to realistic physical
potentials along the lines outlined so far it is necessary to
demonstrate the limited effects variations in the numerical recipe
have and to compare the results obtained with those available by other
means for certain potentials. For this contribution we turn to the
hard-core Yukawa fluid with $z=1.8/\sigma$ for illustration, a system
that has also been considered in a recent study \cite{hrt:14} comparing
various thermodynamically consistent approaches including \HRT\ to
short-ranged potentials.

Among the approximations just mentioned, an important one is the
necessary truncation of eq.~(\ref{eq:matrix:infty}) to a finite number
${N_{\rm cc}}+1$ of basis functions $u_n$ and expansion coefficients
$\g\Q_n$,
$n=0,\ldots,{N_{\rm cc}}$; in fact, not only must ${N_{\rm cc}}$ be finite,
it should
also be rather small if evaluation of the slowly-convergent
$\hat{\cal I}\Q$-integrals at $Q=Q_\infty$ is not to dominate program
execution time. An obvious test for the minimum number of basis
functions to keep is to look at the ${N_{\rm cc}}$-dependence of the phase
behavior predicted as summarized in table~\ref{tb:Ncc}, where the inverse
critical temperature $\beta_c$, the critical density $\rho_c$, and the
coexisting densities at $\beta=0.9/\epsilon$ are recorded; the latter
temperature is sufficiently far away from the critical point so that
the differences in $\rho_v$ and $\rho_l$ are not merely to be
attributed to the differences in $\beta_c$ while maintaining
sufficient separation of the binodal from the boundaries at
${\rho_{\rm min}}=0$ and ${\rho_{\rm max}}$ ({\it v.~i.{}}). From this we
find inclusion of the
core condition to be of vital importance in determining the fluid's
phase behavior, while non-negligible variation of the results remains
even for the highest ${N_{\rm cc}}$-values considered; however, the amount
of
variation especially in $\beta_c$ drops significantly as soon as
${N_{\rm cc}}$ exceeds~5.  ---  The real test for the use of the truncated
eq.~(\ref{eq:matrix:infty})
and the {\it ad hoc} approximation (\ref{eq:dQc2Int:approx}) (easily
demonstrated to be inadequate at least for certain potentials
\cite{ar:th}), is the pair distribution function $g\withQo(r)$ as
obtained from the final values of the expansion coefficients
$\g\withQo_n$ by performing the inverse Fourier transformation implied
by the Ornstein-Zernike equation (\ref{eq:oz}): Employing the \ODE s
following from the consistent application of the decoupling assumption
in order to isolate the approximations pertaining to the
implementation of the core condition from other effects ({\it v.~i.{}}),
the
$g\withQo(r)$ so obtained generally takes on  rather large
values for small $r$ while remaining within a few percent of the
contact value $g\withQo(\sigma+)$ for larger $r$ up to $\sigma-$;
while increasing ${N_{\rm cc}}$ usually does not considerably reduce the
magnitude of $g\withQo(r)$ for $r$ close to~$0$, it instead extends
the $r$-range of rather small $g\withQo(r)$ to ever smaller $r$. At
high density there is no substantial improvement in
$g\withQo(r,\rho)$, $r<\sigma(\rho)$, for ${N_{\rm cc}}>5$, nor for
${N_{\rm cc}}>7$
at low density, a finding corroborated by direct inspection of the
final values of the expansion coefficients $\g\withQo_n$ \cite{ar:th}.
Accordingly, ${N_{\rm cc}}$ should probably be chosen no less than 7
(corresponding to a $6^{\rm th}$ order polynomial in $r$ for $\C\Q(r)$
inside the core), whereas ${N_{\rm cc}}=5$ may still be sufficiently
accurate
for some applications; for ${N_{\rm cc}}<5$, on the other hand, we cannot
expect significantly better results than those from consistently
solving the \PDE\ without implementing the core condition
(cf.~sub-section \ref{sec:ansatz}), which, of course, runs much faster and
at least does not rely on inconsistent approximations. But note that
the core condition is always but poorly met, irrespective of ${N_{\rm
cc}}$,
whenever $f(Q_0,\rho)$ is large (corresponding to the coexistence
region or the critical point's vicinity in implementations relying on
a \PDE). On the other hand, when solving the \PDE\ without
implementing the core condition at all,
$g\withQo(r)$, $r<\sigma$, can, of course, become arbitrarily large;
{\it e.~g.{}}\ for $\beta=0.7/\epsilon$ and $\rho=0.9/\sigma^3$,
$g\withQo(r)=-3.26$ inside the core while the contact value is
$g\withQo(\sigma+)=+1.91$. All in all, while systematic shortcomings
in the pair distribution function $g\withQo(r)$ itself cannot be
avoided in an implementation relying on eqs.~(\ref{eq:matrix:infty}) and
(\ref{eq:dQc2Int:approx}) with finite ${N_{\rm cc}}$, we needs must keep
the core
condition in the calculation due to its bearing on the phase behavior
predicted; this is somewhat at variance with earlier findings
\cite{hrt:10} indicating only a modest influence of the core condition
upon the results, a finding expressly referred to in \cite{hrt:13}.

\begin{figure}
\epsfxsize=8.6cm
\epsfbox{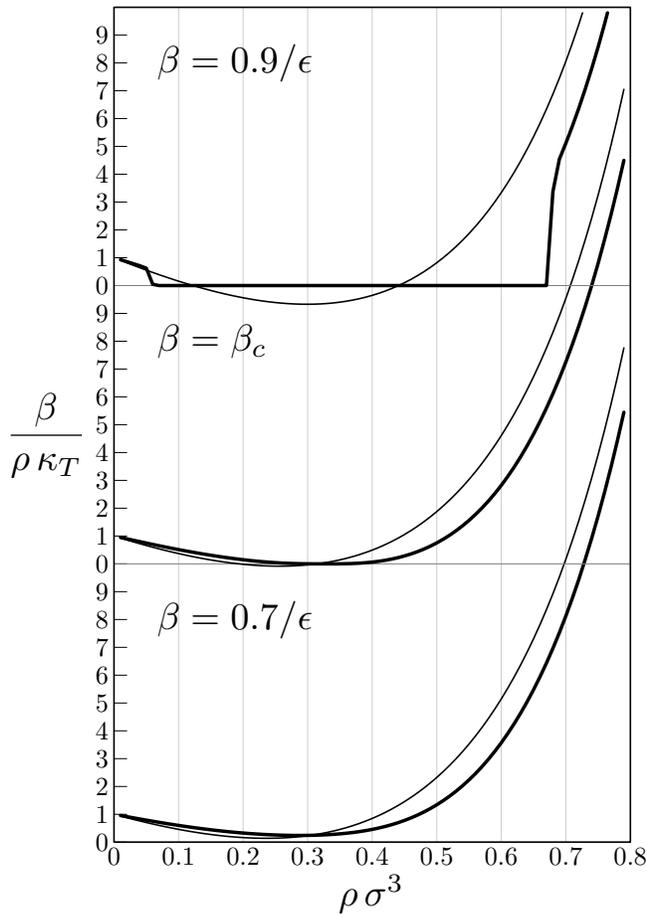}

\caption{Comparison of inverse compressibility of the hard-core
Yukawa system with $z=1.8/\sigma$ as obtained from the \ODE s
following from the decoupling assumption $\alpha\Q(\rho)=0$ (thin
lines) and from a \PDE\ (${N_{\rm cc}}=7$) inconsistently applying this
approximation to the evolution of the core condition expansion
coefficients $\g\Q_n$ only (thick line). In both cases the equations
have been solved on an equispaced density grid with
$\Delta\rho=0.1/\sigma^3$, and the decoupling assumption also served
as a boundary conditon for the \PDE\ at ${\rho_{\rm max}
}=1.0/\sigma^3$. The critical temperature obtained from the
\PDE\ and used in the middle plot is $\beta_c=
0.759497(24)/\epsilon$. }
\label{fig:decoupling}
\end{figure}

On the other hand, as pointed out already in section~\ref{sec:theory},
retaining the core condition is possible only when adopting the
decoupling assumption, {\it viz.{}}\ $\alpha\Q(\rho)=0$, an assumption that
itself suffices to decouple the \PDE\ to a set of \ODE s (which is why
we can use $\alpha\Q({\rho_{\rm max}})=0$ as a boundary condition). But if
such
a procedure is to be considered harmless, the results of the
consistent application of this approximation to the
closure~(\ref{eq:closure}) 
should not differ much from those of the \PDE\ applying the decoupling
assumption only in order to get rid of the second $\hat{\cal I}$-integral
in
eq.~(\ref{eq:matrix:infty}). However, the calculations summarized in
fig.~\ref{fig:decoupling} clearly show that the two approaches yield very
different results so that we cannot rule out a non-negligible effect
on the structural and thermodynamic properties predicted: Most
importantly, the \ODE s cannot reproduce well-defined phase
boundaries, and they even yield slightly negative inverse
compressibility $1/\kappa_T$ in what would otherwise be the
coexistence region. Preserving the structure of the \PDE, so that
thermodynamic consistency is at least partly implemented by the \PDE's
coefficients $d_{0i}$ as defined in eq.~(\ref{eq:d0i}), seems sufficient to
remedy these deficiencies; at any rate, we have to accept the
decoupling assumption as indispensable for the implementation of the
core condition.

Let us now turn to a brief discussion of several other aspects of
\HRT\ in its present formulation and the potential problems they may
present; as most of these effects are much more prominent for the
square-well and multistep potentials defined in sub-section
\ref{sec:potential}, we relegate more thorough treatment to
\cite{ar:sw,ar:th}. --- Clearly, the importance of retaining the
structure of a \PDE\ mandates closer examination of the properties of
the set of densities, especially since the terms corresponding to the
finite difference approximation to the operator
$(\partial^2/\partial\rho^2)$ have been found the primary limiting
factor for the quality of the numerical solution (cf.~the definition
of $\epsNum$ in sub-section~\ref{sec:solver}); while obvious for
subcritical temperatures at low $Q$ where the near-discontinuity of
$f$ at the phase boundary betrays the smoothness assumptions
underlying finite difference schemes, this is true even for rather
small $f(Q,\rho)$. For our hard-core Yukawa system, however, the
results' stability with respect to a variation of the density grid or
the location and nature of the high density boundary
(cf.~table~\ref{tb:ansatz}) is rather satisfactory as long as ${\rho_{\rm
max}}-\rho_l$ exceeds
several $\Delta\rho$: just as expected from fig.~\ref{fig:decoupling}, the
\ODE\ used at the boundary forces the corresponding density to lie
outside the coexistence region, which carries over to nearby densities
by virtue of the \PDE's discretization. Despite the identical phase
behavior found for different boundary conditions as evidenced by
table~\ref{tb:ansatz}, the question of which condition to impose at
${\rho_{\rm max}}$
is far from irrelevant; indeed, when imposing the
\LOGA/\ORPA-condition $\g\Q_0=0$ at ${\rho_{\rm max}}$ without making use
of
decoupling at all, inspection of the solution generated close to
${\rho_{\rm max}}$ clearly shows the need to replace the original
implementation's approach by an ansatz less inconsistent. ---
Similarly, the stiffness of the \PDE\ for subcritical temperatures
shortly touched upon in section \ref{sec:theory} is easily detected by
employing the criteria discussed in sub-section~\ref{sec:solver}; still,
for $v^{{\rm hcy}}$ with $z=1.8/\sigma$ and ${N_{\rm cc}}=7$, using
pre-determined
step sizes and resigning on any control of the predictor-corrector
scheme's convergence we obtain an approximate solution at $Q_0$
sufficiently stable outside the coexistence region; thus, while the
solution generated in the region of large $f(Q,\rho)$ necessarily
differs from the true solution in a fundamental way, the influence of
which on the data produced outside the coexistence region is not
assessed easily, stiffness here appears less
pressing a concern than for other systems \cite{ar:sw}.

In addition to the internal consistency of the results, we also have
to make contact with data available by other means; as this paper's
purpose is presentation of our software and discussion of some general
aspects of \HRT\ rather than a comprehensive study of the hard-core
Yukawa system, we here restrict ourselves to the data of table~I of
\cite{hrt:14} containing the predictions of various thermodynamically
self-consistent liquid state theories including \HRT\ in the original
implementation, which are found to yield $T_c$ ranging from
$1.193\epsilon/\kB$ to $1.219\epsilon/\kB$ ($\beta_c=1/\kB\,T_c$
between $0.820/\epsilon$ and $0.838/\epsilon$), as well as the Monte
Carlo (\MC) result $T_c=1.212(2)\epsilon/\kB$
($\beta_c=0.825(2)/\epsilon$). As is apparent from our table~\ref{tb:Ncc},
our implementation's predictions for $1\le{N_{\rm cc}}\le4$ fall precisely
into this range and, for ${N_{\rm cc}}\in\{3,4\}$, are well compatible with
simulation results; that very same ${N_{\rm cc}}$-range, on the other hand,
we
have seen ({\it v.~s.{}}) is characterized by gross violation of the core
condition due to an insufficient number of basis functions retained in
the truncated eq.~(\ref{eq:matrix:infty}). As we further increase ${N_{\rm
cc}}$ so
that the core condition is obeyed to a certain extent ({\it v.~s.{}}),
however,
$\beta_c$ drops dramatically to values far outside the range quoted in
\cite{hrt:14}; while the trend of increasing $\beta_c$ evident for
${N_{\rm cc}}\ge7$ indicates that \HRT\ might match the \MC\ predictions
for
${N_{\rm cc}}\sim15$, we have not performed these \CPU\ intensive
calculations.

With this we conclude our superficial sketch of the software we have
written and the appraisal of the results it typically produces as
illustrated for the hard-core Yukawa potential $v^{{\rm hcy}}$ with
$z=1.8/\sigma$: it should be apparent that \HRT\ in its current
formulation, while presenting substantial difficulties discussed here
as well as in \cite{ar:sw,ar:th}, is well capable of predicting
structural and thermodynamic properties of simple one-component
fluids; at the same time, the computational difficulties mentioned and
the approximations introduced to render the numerics tractable cannot
always be shown to be harmless so that great care has to be exercised
in the interpretation of isolated results. The fully modular design of
our program and the high degree of flexibility brought about by the
adoption of a meta language and code construction techniques are key
factors in facilitating additional evaluations, providing a means of
separating different approximations' effects; at the same time, our
implementation makes for a versatile tool for the systematic
exploration of \HRT\ for one-component fluids in its present or
alternative formulations.

\section{Acknowledgments}

{\sc ar} thanks G.~Stell (Stony Brook, N.Y.) and G.~Pastore (Trieste)
for stimulating discussions, as well as D.~Pini (now in Milan) for
sharing his experiences with HRT and the code of the original HRT
program. 
The authors gratefully acknowledge support from {\it\"Osterreichischer
Forschungsfonds} under project number P13062-TPH.

\appendix

\section{Re-formulation in not-quite quasi-linear form}

\label{sec:app:rewrite}

As noted in section~\ref{sec:theory}, it is advantageous to replace the
highly non-linear \PDE\ in the modified free energy $\A\Q$ implied by
eqs.~(\ref{eq:dQA}), (\ref{eq:consistency}), and the core condition, with a
formulation akin to a quasi-linear one. In complete analogy to
\cite{hrt:10}, we define the auxiliary function $f(Q,\rho)$ via
\begin{displaymath}
\ln\left(1-{\tilde\phi(Q,\rho)\over\tilde\C\Q(Q,\rho)}\right)
=
f(Q,\rho)\,\tilde u_0^2(Q,\rho) -
{\tilde\phi(Q,\rho)\over\tilde\K\Q(Q,\rho)}\,;
\end{displaymath}
in the ideal gas limit, {\it i.~e.{}}\ for $\rho\to0$, we immediately find
from
the divergence of the terms $-1/\rho$ in $\tilde c\Ref_2$ and $\tilde\C\Q$
that $f(Q,0)$ = $\partial f(Q,0)/\partial\rho = 0$, which is a
convenient boundary condition most implementations rely on. In the
above definition we have taken advantage of some freedom regarding the
$f$-term to reduce the number of floating-point multiplications; as
$u_0$ is usually chosen to be strictly proportional to $\phi$, our $f$
differs only by a constant factor from the choice of \cite{hrt:10} which
can be recovered by replacing the factor $\tilde u_0^2(Q,\rho)$ by
$\tilde\phi^2(Q,\rho)$; note, however, that the expressions we give here
remain valid for a slightly more general choice of $u_0$, allowing the
proportionality of the basis function and the perturbational part of
the potential to hold only up to order $\tilde\phi^2$ --- a freedom that
might be exploited to use more appropriate basis functions, giving
$\C\Q$ a larger range in $r$-space. Also we should point out that the
condition of non-singular coefficients in the \PDE\ (\ref{eq:quasilinear}),
the very reason for the restriction on the relation between $u_0$ and
$\phi$ just mentioned as well as for the introduction of the term
involving $\tilde\phi/\tilde\K\Q$, only fixes a minimum exponent for the
$\tilde u_0$-factor with which to multiply $f$ \cite{ar:th}.

Inserting the above definition for $f$ into the relevant equations of
section~\ref{sec:theory} and eliminating the expansion coefficient
$\g\Q_0$ via the consistency condition (\ref{eq:consistency}), we can
easily
re-cast the original \PDE\ in the form of eq.~(\ref{eq:quasilinear}).
Dropping the obvious arguments and with the shorthand
notations
\begin{displaymath}
\begin{array}{c}
\eps = 1-{\tilde\phi\over\tilde\C\Q}
= e^{f\,\u_0^2 + x_\phi}\qquad
\epsbar=\eps-1
\\
x_\phi=-\tilde\phi/\tilde\K\Q\qquad
\tilde\phi_0=\tilde\phi(0,\rho)\qquad
\tilde\G_0=\tilde\G\Q(0,\rho)
\end{array}
\end{displaymath}
the coefficients of eq.~(\ref{eq:quasilinear}) can be written as
\cite{ar:th}

\begin{equation}\label{eq:d0i}
\begin{array}{rl}
d_{00}={}
&+{\partial\tilde\phi\over\partial Q}
\,\left({\tilde\phi_0^2\over\tilde\K\Q\,\tilde\phi^2}
- {\tilde\K\Q\,\epsbar^2\,\tilde\phi_0^2\over\eps\,\tilde\phi^4}
- {2 f\over\tilde\phi}\right)\\
&+ {\partial\tilde\K\Q\over\partial Q}
\,\left({\epsbar^2\,\tilde\phi_0^2\over\eps\,\tilde\phi^3}
- {\tilde\phi_0^2\over\left(\tilde\K\Q\right)^2\,\tilde\phi}\right)\\
&- {\partial^2\u_0^2\over\partial\rho^2}
\,{Q^2\,\epsbar^2\,f\,\tilde\phi_0\over4\pi^2\,\eps\,\tilde\phi^2}\\
&- {\partial^2x_\phi\over\partial\rho^2}
\,{Q^2\,\epsbar^2\,\tilde\phi_0\over4\pi^2\,\eps\,\tilde\phi^2}\\
&- {\partial\tilde\G\Q_0\over\partial Q}
\,{\epsbar^2\,\tilde\phi_0\,\over\eps\,\tilde\phi^2}\\
d_{01}={}
&-{\partial\u_0\over\partial\rho}
\,{Q^2\,\epsbar^2\over\pi^2\,\eps\,\tilde\phi}\\
d_{02}={}
&-{Q^2\,\epsbar^2\over4\pi^2\,\eps\,\tilde\phi_0}.
\end{array}
\end{equation}
When evaluating these expressions care has to be taken whenever
$\tilde\phi(Q,\rho)$ is close to zero as all terms of orders
$1/\tilde\phi^2(Q,\rho)$ and $1/\tilde\phi(Q,\rho)$ must cancel; noting
that $(\epsbar^2\,\tilde\phi_0^2/\eps\tilde\phi^3) -
(\tilde\phi_0^2/(\tilde\K\Q)^2\,\tilde\phi)$, the coefficient of
${\partial\tilde\K\Q/\partial Q}$ in $d_{00}$, can be written in terms
of the $(\partial\phi/\partial Q)$-coefficient, in our numerical work
the calculation of the terms affected by the cancellation  for small
$\tilde\phi(Q,\rho)$ proceeds via application of a
fifth-order Taylor expansion of $(\tilde\phi_0^2/\tilde\K\Q\,\tilde\phi^2)
-
(\tilde\K\Q\epsbar^2\,\tilde\phi_0^2/\eps\,\tilde\phi^4) - (2
f/\tilde\phi)$;
one should note that, even though the criterion for switching between
the full analytic expressions and said expansion depends on $\epsNum$,
the order of this expansion (and a similar one for $\epsbar/\tilde\phi$)
is not increased for very small values of $\epsNum$, which is one of
the few hard-coded limitations of our program.

From the given expressions for the $d_{0i}$ two more aspects are
obvious: (i) there is substantial further simplification for a
$\rho$-independent potential (as $u_0$ depends upon the density only
through the perturbational part $\phi$ of the potential, and the basis
functions entering $\G\Q$ only through the reference system's
hard-core radius $\sigma(\rho)$), and (ii) the \PDE\ does not fall into the
class of quasi-linear \PDE{}s due to the presence of the second
$\rho$-derivative of $x_\phi$ in the coefficient $d_{00}$.

\section{Basis functions for the core condition}

\label{sec:app:basis}

While the basis function $u_0(r,\rho)\propto w(r,\rho)$ in the closure
(\ref{eq:closure}) is fixed by the potential used, in principle there is
ample freedom in choosing the $u_n(r)$, $n\ge1$. Of course, when
truncating eq.~(\ref{eq:matrix:infty}) after $1+{N_{\rm cc}}$ basis
functions it is
natural to ask for the set $\{u_1(r),\ldots,u_{N_{\rm cc}}(r)\}$ to span
the
space of polynomials of order up to ${N_{\rm cc}}-1$, so that $u_n(r)$ will
generally be a polynomial or order $n-1$ in $r$; but whereas different
polynomials of this type do not alter the function space, their choice
has implications for the numerical properties of the matrix equations
implementing the core condition as well as, to a certain extent, for
the convergence of the $\hat{\cal I}$-integrals to be evaluated at
$Q=Q_\infty$ (cf.\ sub-section~\ref{sec:ansatz}). Other than the original
implementation that relied on an affine transformation of the Legendre
polynomials, we found it convenient to  choose
$u_n(r,\rho)$ simply proportional to a power of $r$ and normalized to
$\tilde u_n(0,\rho)=1$.  Thus, dropping the obvious argument $\rho$, we
have
\begin{displaymath}
u_n(r)
= \left\{
\begin{array}{ccc}
{n+2\over4\pi\sigma^3}\,\left({r\over\sigma}\right)^{n-1}&:&r<\sigma\\
0&:&r>\sigma\,,
\end{array}
\right.
\end{displaymath}
the Fourier transform of which is
\begin{displaymath}
\tilde u_n(k)
= {n+2\over(\sigma k)^{n+2}}\,
\left[
n!\,\cos{n\pi\over2}
-\sum_{j=0}^n {n!\over(n-j)!}\,(\sigma k)^{n-j}\,
\cos\left(\sigma k+{j\pi\over2}\right)
\right],
\end{displaymath}
an expression used for $\sigma k>n$ only for numerical reasons. For
smaller $k$ we rely on the expansion
\begin{displaymath}
\tilde u_n(k)
= (n+2)\,
\sum_{j=0}^\infty {(-1)^j\over(n+2j+2)\,(2j+1)!}\,(\sigma k)^{2j}\,,
\end{displaymath}
truncating the series after $N_n$ terms, where $N_n$ is the smallest
number such that
\begin{displaymath}
{n+2\over n+2N_n+4}\,{n^{2N_n+2}\over(2N_n+3)!}
\le \epsNum\, p^{[u_n]}_{N_n}
\end{displaymath}
with a customization factor $p^{[u_n]}_{N_n}$ of order unity.

\begin{table}
\begin{tabular}{|c|ccccc|}
  ${N_{\rm cc}}$&$\beta_c\,\epsilon$&
     $\rho_v(\beta^*=0.9/\epsilon)\,\sigma^3$&
     $\rho_l(\beta^*=0.9/\epsilon)\,\sigma^3$&
$\rho_c\,\sigma^3$&\\
\hline
--&0.83164(39)&0.115(5)&0.565(5)&0.325(30)&\\
\hline
1&0.82070(39)&0.105(5)&0.575(5)&0.315(30)&\\
2&0.82148(39)&0.105(5)&0.565(5)&0.315(10)&\\
3&0.82227(39)&0.105(5)&0.565(5)&0.315(30)&\\
4&0.82227(39)&0.105(5)&0.565(5)&0.315(30)&\\
5&0.77676(20)&0.075(5)&0.645(5)&0.320(15)&\\
6&0.75527(20)&0.055(5)&0.685(5)&0.325(30)&\\
7&0.75957(20)&0.065(5)&0.675(5)&0.330(25)&\\
8&0.77305(39)&0.065(5)&0.645(5)&0.320(35)&\\
9&0.78555(39)&0.075(5)&0.615(5)&0.315(20)&\\
\end{tabular}
\caption{Dependence of inverse critical temperature
$\beta_c=1/\kB\,T_c$, coexisting densities $\rho_v$ and $\rho_l$ at
$\beta=0.9/\epsilon$, and critical density $\rho_c$ for a hard-core
Yukawa potential with $z=1.8/\sigma$ on the number of basis
functions. The results reported have been obtained from \PDE s
retaining ${N_{\rm cc}}+1$ basis functions or (first line) not implementing
the core condition at all, with $\epsNum=10^{-2}$ and
${\rho_{\rm max}}=1.0/\sigma^3$. We have checked that the differences
summarized here cannot be explained by the ${N_{\rm cc}}$-dependence of the
upper integration limits in evaluating the
$\hat{\cal I}\withQ{Q_\infty}$-integrals (cf.\
sub-section~\ref{sec:ansatz}). }
\label{tb:Ncc}
\end{table}

\begin{table}
\begin{tabular}{|ccc|ccccc|}
  method&${N_{\rm cc}}$&bound.~cond.&
    $\beta_c\,\epsilon$&
    $\rho_v(\beta=0.9/\epsilon)\,\sigma^3$&
    $\rho_l(\beta=0.9/\epsilon)\,\sigma^3$&
$\rho_c\,\sigma^3$&\\
\hline
  no core condition&--&$\g\Q_0({\rho_{\rm max}})=0$&
0.831573(91)&0.115(5)&0.565(5)&0.330(20)&\\
\hline
  decoupling&7&$\g\Q_0({\rho_{\rm max}})=0$&
0.759429(91)&0.055(5)&0.675(5)&0.325(15)&\\
  decoupling&7&$\alpha\Q({\rho_{\rm max}})=0$&
0.759429(91)&0.055(5)&0.675(5)&0.325(15)&\\
\end{tabular}
\caption{Inverse critical temperature $\beta_c=1/\kB\,T_c$,
coexisting densities $\rho_v$ and $\rho_l$ at $\beta=0.9/\epsilon$,
and critical density $\rho_c$ for a hard-core Yukawa potential with
$z=1.8/\sigma$ as predicted by various combinations of approximations
and boundary conditions at ${\rho_{\rm max}}=1/\sigma^3$. Again, the
results
reported have been obtained from \PDE s retaining ${N_{\rm cc}}+1$ basis
functions or (first line) not implementing the core condition at all,
with $\epsNum=10^{-2}$ and ${\rho_{\rm max}}=1.0/\sigma^3$.}
\label{tb:ansatz}
\end{table}

\begin{thebibliography}{99}
\label{sec:biblio}

\bibitem{b:hrt:1}A.~Parola, L.~Reatto, Adv.~Phys.\ {\bf44} (1995) 211.

\bibitem{hrt:1}A.~Parola, L.~Reatto, Phys.~Rev.~Lett.\ {\bf53} (1984) 2417.

\bibitem{hrt:2}A.~Parola, L.~Reatto, Phys.~Rev.~A {\bf31} (1985) 3309.

\bibitem{hrt:8}A.~Parola, D.~Pini, L.~Reatto, Phys.~Rev.~E\ {\bf48}
(1993) 3321.

\bibitem{hrt:12}A.~Parola, J.~Phys.~C: Solid State Phys.~{\bf19}
(1986) 5071.

\bibitem{hrt:4}A.~Meroni, A.~Parola, L.~Reatto, Phys.~Rev.~A {\bf42}
(1990) 6104.

\bibitem{hrt:7}A.~Meroni, L.~Reatto, M.~Tau, Mol.~Phys.\ {\bf80}
(1993) 977.

\bibitem{hrt:10}M.~Tau, A.~Parola, D.~Pini, L.~Reatto, Phys.~Rev.~E
{\bf52} (1995) 2644.

\bibitem{hrt:14}C.~Caccamo, G.~Pellicane, D.~Costa, D.~Pini, G.~Stell,
Phys.~Rev.~E {\bf60} (1999) 5533.

\bibitem{hrt:17}L.~Reatto, A.~Parola, J.~Phys.:\ Condens.\ Matter
{\bf8} (1996) 9221.

\bibitem{hrt:18}F.~Barocchi, P.~Chieux, R.~Fontana, R.~Magli,
A.~Meroni, A.~Parola, L.~Reatto, M.~Tau, J.~Phys.:\ Condens.\ Matter
{\bf9} (1997) 8849.

\bibitem{ar:sw}A.~Reiner, G.~Kahl, to be published.

\bibitem{hrt:11}L.~Reatto, Phys.~Lett.\ {\bf72A} (1979) 120.

\bibitem{wca:1}D.~Chandler, J.~D.\ Weeks, Phys.~Rev.~Lett.\ {\bf25}
(1970) 149.

\bibitem{wca:2}D.~Chandler, J.~D.\ Weeks, H.~C.\ Andersen, J.\ Chem.\
Phys.\ {\bf54} (1971) 5237.

\bibitem{wca:3}H.~C.\ Andersen, J.~D.\ Weeks, D.~Chandler, Phys.~Rev.~A
{\bf4} (1971) 1597.

\bibitem{hrt:3}A.~Parola, A.~Meroni, L.~Reatto, Phys.~Rev.~Lett.\
{\bf62} (1989) 2981.

\bibitem{oz}L.~S.~Ornstein, F.~Zernike, Proc.\ Akad.\ Sci.\ (Amsterdam)
{\bf 17} (1914) 793.

\bibitem{loga1}G.~Stell, Phys.~Rev.~{\bf184} (1969) 135.

\bibitem{loga2}G.~Stell, J.\ Chem.\ Phys.\ {\bf55} (1971) 1485.

\bibitem{orpa}H.~C.~Andersen, D.~Chandler, J.\ Chem.\ Phys.\ {\bf55}
(1971) 1497.

\bibitem{ar:th}A.~Reiner, PhD Thesis, Technische Universit\"at Wien,
2002. Available on the world wide web from {\tt
http://purl.oclc.org/NET/a-reiner/dr-thesis/}.

\bibitem{hs:8}J.~L.~Lebowitz, Phys.~Rev.~{\bf133} (1964) A895.

\bibitem{hs:2}D.~Henderson, E.~W.~Grundke, J.\ Chem.\ Phys.\ {\bf63}
(1975) 601.

\bibitem{numpde:cfl}R.~Courant, K.~O.\ Friedrichs, H.~Lewy,
Mathematische Annalen {\bf100} (1928) 32.

\bibitem{hrt:13}D.~Pini, A.~Parola, L.~Reatto, J.\ Stat.\ Phys.\
{\bf100} (2000) 13.

\end{thebibliography}
\end{document}